\documentclass[10pt]{iopart}

\usepackage{iopams}
\usepackage{color}
\usepackage{amsmath}
\usepackage{amssymb}
\usepackage{graphicx}
\usepackage{cite}

\begin{document}

\title{Role of the radial electric field in the confinement of energetic ions in the Wendelstein 7-X stellarator}

\author{M. Arranz, J. L. Velasco, I. Calvo and D. Carralero}

\address{Laboratorio Nacional de Fusión, CIEMAT, 28040 Madrid, Spain}
\ead{marcos.arranz@ciemat.es}
\vspace{10pt}
\begin{indented}
\item Feb 2026
\end{indented}

\begin{abstract}
Good fast-ion confinement is an essential requirement for a fusion reactor. The magnetic configuration of the Wendelstein 7-X (W7-X) stellarator is partially optimized in this regard in a reactor-relevant scenario: it is expected to show improved fast-ion confinement when $\beta$ is high and the effect of the radial electric field is negligible. The experimental validation of this optimization is difficult since, with the available power, achieving high $\beta$ under appropriate conditions for the validation is challenging and the effect of the radial electric field is inevitable. In this work, the confinement of fast ions in W7-X has been studied numerically for a variety of scenarios via the ASCOT5 code. The effect of the radial electric field on fast-ion losses is confirmed to be equivalent to the one produced by $\beta$, and this is characterized by means of scans on both parameters. Through a preliminary study with experimentally-based profiles, a viable scenario is identified that takes advantage of this effect for the experimental validation of the optimization strategy of W7-X.
\end{abstract}

\vspace{2mm}

\ioptwocol

\markboth{Role of the radial electric field in the confinement of energetic ions in W7-X}{Role of the radial electric field in the confinement of energetic ions in W7-X}

\section{Introduction}
\label{introduction}

Stellarators are a viable alternative to tokamaks in the development of a magnetic confinement fusion reactor, since, among other inherent advantages, they are intrinsically able to operate in steady state and are free from disruptions \cite{helander2012stellarator}. However, stellarators also present challenges, such as their complex and intricate geometry due to their necessary lack of axial symmetry \cite{helander2014theory}. Consequently, generic stellarators exhibit bad neoclassical confinement (the one related to inhomogeneities in the magnetic field and collisions between particles) compared to tokamaks. Fortunately, the greater number of degrees of freedom in a stellarator (associated with its three-dimensional character) allows for an optimization of its magnetic configuration that reduces these disadvantages \cite{calvo2014optimizing,plunk2019direct,landreman2022magnetic,mata2022direct,landreman2022mapping,sanchez2023quasi,goodman2023constructing,dudt2024magnetic,garcia2024reduced}, reaching tokamak-like confinement.

In particular, proper confinement of fusion-generated alpha particles is essential for a fusion reactor. The most restrictive requirement is imposed by the heat load on the reactor walls: alpha particles that are rapidly lost, and thus retain most of their initial energy, could potentially damage the reactor components exposed to the plasma. Moreover, these alpha particles are intended to contribute to heat the plasma. Therefore, their confinement time must be comparable to or larger than the time it takes them to transfer their energy to the plasma and thermalize.

Regarding energetic ions, the non-axisymmetric geometry of stellarators causes trapped-particle orbits to be unconfined. Therefore, in a generic stellarator, energetic ions leave the plasma in a brief period of time, something incompatible with a reactor operation. This can be solved through the previously mentioned optimization of the magnetic configuration. If the bounce-average radial drift is null for all trapped particles, their orbits are confined, implying low neoclassical transport. This is the case in omnigenous \cite{hall1975three,cary1997helical,parra2015less} and piecewise omnigenous \cite{velasco2024piecewise,velasco2025exploration,calvo2025new} configurations. Quasi-isodynamic (QI) configurations are a specific family of omnigeneus stellarators where contours of constant magnetic field strength close poloidally.

In current fusion devices, fusion reactions are not occurring (or at least not to a significant extent), and, therefore, they lack fusion-generated alpha particles. Instead, energetic ions are provided via external heating schemes, such as Neutral Beam Injection (NBI) or Ion Cyclotron Resonance Heating (ICRH) \cite{menon1981neutral,stork1991neutral,kirov2020synergistic,rust2011w7xnbi,machielsen2023fast}. These heating schemes generate the so-called fast ions: ions with energies considerably higher than the bulk plasma ion temperature ($\sim 10 - 100$ times higher). These fast ions may be ions from the bulk plasma that have been accelerated to high energies (as in the case of ICRH), or ions introduced into the plasma by the heating scheme (as in the case of NBI). Despite not being fusion-generated alpha particles, these fast ions should exhibit dynamics similar to that of fusion-generated alpha particles from a reactor, since they present the same normalized Larmor radius. Hence, these fast ions are employed to study the behavior of future fusion-generated alpha particles prior to building a reactor. Nonetheless, the ratio of their energy to that of the plasma is lower than in a reactor, which, as we will see, complicates such studies.

Wendelstein 7-X (W7-X) is an example of a partially optimized stellarator with respect to, among other criteria, the confinement of energetic ions \cite{nuhrenberg1995overview,klinger2016performance,bosch2017final}. Specifically, W7-X was optimized to be close to quasi-isodynamicity. This optimization has led to record performance \cite{wolf2017major,klinger2019overview}, thanks to low neoclassical transport \cite{beidler2021demonstration}. As we will discuss in more detail in section \ref{theory}, in W7-X, good fast-ion confinement is expected at high $\beta$ (compared to low $\beta$) \cite{drevlak2014fast} as a result of the combination of three effects: (i) the orbit-averaged radial drift of trapped particles is small, (ii) the diamagnetic effect causes the radial variation of the magnetic field strength to increase at high $\beta$ (see e.g. \cite{herbemont2022finite}) and (iii) the poloidal precession caused by this radial variation overcomes the orbit-averated radial drift, which improves fast-ion confinement \cite{wobig1999theory,velasco2021model}.

The small magnitude of the orbit-averaged radial drift has already been experimentally verified through its effect on bulk energy transport in \cite{beidler2021demonstration}. The increased radial variation of the magnetic field strength due to the diamagnetic effect has also been indirectly observed in \cite{wolf2018electron}. However, (iii) still needs to be validated.

Ideally, one would want to perform a scan on $\beta$ in W7-X and measure the expected enhancement in fast-ion confinement. Nevertheless, reaching high $\beta$ (specially under the appropriate conditions for a validation exercise) is challenging \cite{wolf2017major,klinger2019overview}. In addition, radial electric fields of the size of those measured \cite{pablant2018core,pablant2020investigation,carralero2020characterization,estrada2021radial,carralero2021experimental,alonso2022plasma} are known to affect the poloidal precession of fast ions \cite{kolesnichenko2006effects}. Specifically, when the radial electric field from the experiment is taken into account, it has the largest contribution to the poloidal drift (even at high $\beta$) and, therefore, determines fast-ion confinement \cite{green2026energetic}. Hence, the effect produced by a change in $\beta$ could be concealed by a change in the radial electric field. In fact, previous works \cite{lazerson2021modeling,faustin2016fast} point to the difficulty, due to the effect of the electric field, of studying the configuration dependence of fast-ion confinement in W7-X.

In this work, instead of as a nuisance, we regard the fine dependence of fast-ion confinement on the radial electric field as an opportunity for an experimental validation of W7-X optimization strategy. Namely, to validate the improvement of fast-ion confinement due to an increase in the poloidal precession. A primary objective of this work is to check that the radial electric field produces an analogous effect on fast-ion confinement to $\beta$, for achievable values of the electric field. Thereafter, this work aims to assess to what extent a scan in radial electric field can actually be performed in W7-X and used for an experimental validation of its optimization.

In this validation exercise, it is crucial to be as quantitatively precise as possible because QI magnetic configurations are typically optimized to exhibit good fast-ion confinement above a certain $\beta$. An inaccurate prediction of the role of the poloidal drift could shift the scenario of good fast-ion confinement towards higher $\beta$ values than desirable, potentially exceeding the operational limits of a reactor.

The rest of the paper is organized as follows. In section \ref{theory}, the coordinates employed in the paper are defined and some theoretical concepts are introduced. In section \ref{workflow}, the workflow and the methodology for all the simulations and calculations of this work are described. In section \ref{academic_scan}, we show and discuss the results from an ideal scan with academic profiles, enabling the effect of the radial electric field and $\beta$ to be disentangled. Afterwards, we assess the possibility of performing an equivalent scan in the experiment and, in order to do so, results for an experimentally-based scan are shown and discussed in section \ref{experiment_scan}. Finally, conclusions are outlined in section \ref{conclusions}.

\section{Coordinates and theoretical background}
\label{theory}

Let us assume strongly magnetized fast ions, so that $\rho_* \equiv \rho / L \ll 1$, where $\rho = v_{\perp} / \Omega$ is the fast-ion Larmor radius or gyroradius and $L$ is a characteristic macroscopic length of the order of the stellarator minor radius. Here, $\Omega = Z e B / m$ is the fast-ion gyrofrequency, $Z e$ and $m$ are the fast-ion charge and mass respectively, $B = \left| \mathbf{B} \right|$ is the magnetic field strength, $\mathbf{v}$ is the fast-ion velocity, $v = \left| \mathbf{v} \right|$,
\begin{equation} \label{eq_def_v_perp}
    v_{\perp} = \sqrt{v^2-v_{\|}^2}
\end{equation}
is the velocity perpendicular to the magnetic field,
\begin{equation} \label{eq_def_v_par}
    v_{\|} = \mathbf{v} \cdot \mathbf{b}
\end{equation}
is the velocity parallel to the magnetic field and $\mathbf{b} = \mathbf{B} / B$. The smallness of $\rho_*$ allows us to calculate fast-ion transport in the framework of drift-kinetics \cite{hazeltine1973plasma,herbemont2022finite}, which describes the motion of the guiding-centers of charged particles in strongly magnetized plasmas.

We will use $s$, $\alpha$ and $l$ as spatial coordinates. Here,
\begin{equation}
    s = \frac{\Psi}{\Psi_{LCFS}}
\end{equation}
is a radial variable that labels the flux surfaces, where $2 \hspace{0.02 cm} \pi \Psi$ is the toroidal flux through the flux surface and $2 \hspace{0.02 cm} \pi \Psi_{LCFS}$ is the toroidal flux through the Last Closed Flux Surface (LCFS). The coordinate
\begin{equation}
    \alpha = \theta - \iota \zeta
\end{equation}
is a poloidal angle that labels field lines on each flux surface, where $\theta$ and $\zeta$ are poloidal and toroidal Boozer angles and $\iota$ is the rotational transform. Finally, $l$ is the arc-length of the magnetic field line. In this set of coordinates, the magnetic field, $\mathbf{B}$, takes the form
\begin{equation}
    \mathbf{B} = \Psi_{LCFS} \hspace{0.05 cm} \boldsymbol{\nabla} s \times \boldsymbol{\nabla} \alpha .
\end{equation}

In velocity space, we will use coordinates $\mathcal{E}$, $\mu$ and $\sigma$, where
\begin{equation} \label{eq_def_energy}
    \mathcal{E} = \frac{1}{2} m v^2 + Z e \Phi
\end{equation}
is the fast-ion energy,
\begin{equation} \label{eq_def_mu}
    \mu = \frac{m v_{\perp}^2}{2 B}
\end{equation}
is the fast-ion magnetic moment and
\begin{equation} \label{eq_def_sigma}
    \sigma = \frac{v_{\|}}{\left| v_{\|} \right|} = \pm 1
\end{equation}
is the sign of the parallel velocity. Here, $\Phi$ is the electrostatic potential, which, in this work, is assumed to depend only on $s$.

In the absence of collisions, the fast-ion energy and the fast-ion magnetic moment are conserved quantities (being $\mu$ only an adiabatic invariant, while $\mathcal{E}$ is an exact invariant). Combining equations (\ref{eq_def_energy}), (\ref{eq_def_mu}) and (\ref{eq_def_v_perp}), $v_{\|}$ can be related to $\mathcal{E}$ and $\mu$,
\begin{equation}  \label{eq_v_par_expression}
    v_{\|} = \sigma \sqrt{\frac{2 \left( \mathcal{E} - \mu B - Z e \Phi \right)}{m}} .
\end{equation}

The guiding-center velocity $\dot{\mathbf{R}}$ can be written as the sum of the velocity parallel to the magnetic field and a perpendicular drift velocity $\mathbf{v}_d$,
\begin{equation} \label{eq_gc_velocity}
    \dot{\mathbf{R}} = v_{\|} \mathbf{b} + \mathbf{v}_d ,
\end{equation}
where $\left| \mathbf{v}_d / v_{\|} \right| \sim \rho_*$. The drift velocity $\mathbf{v}_d = \mathbf{v}_M + \mathbf{v}_E$ is the sum of the magnetic drift
\begin{equation} \label{eq_def_vM}
    \mathbf{v}_M = \frac{2 \mathcal{E}}{Z e} \left( 1 - \frac{\lambda B}{2} \right) \frac{\mathbf{B} \times \boldsymbol{\nabla} B}{B^3}
\end{equation}
and the $E \times B$ drift
\begin{equation} \label{eq_def_vE}
    \mathbf{v}_E = \frac{\mathbf{E} \times \mathbf{B}}{B^2} ,
\end{equation}
where $\lambda \equiv \mu / \mathcal{E}$ is the \textit{pitch-angle} variable and $\mathbf{E} = - \boldsymbol{\nabla} \Phi$ is the electric field.

Given that, to lowest order in $\rho_*$, guiding-centers move along the magnetic field lines, we will distinguish between two types of particle trajectories. Additionally, since the time scale of motion along the field line is much faster than the motion perpendicular to it, we will average quantities along this motion.

If $\lambda B_{max} < 1 - Z e \Phi / \mathcal{E}$, where $B_{max}$ is the maximum of $B$ on the surface, $v_{\|}$ never vanishes and, therefore, the particle always moves in the same direction with respect to $\mathbf{B}$; being able to circulate around the torus and cover the entire flux surface (on ergodic surfaces). These particles are called \textit{passing} particles. On the other hand, if $\lambda B_{max} \geq 1 - Z e \Phi / \mathcal{E}$, the particle will move in a given direction along $\mathbf{B}$ until it reaches a point where $v_{\|}$ vanishes and reverses its sign. Thus, the particle will be trapped bouncing back and forth along the magnetic field line between the points where $v_{\|} = 0$, the so-called \textit{bounce points}. These particles are called \textit{trapped} particles.

To next order in $\rho_*$, guiding-centers drift across magnetic field lines. For passing particles, one can prove that the net radial displacement, caused by the term $\mathbf{v}_d \cdot \boldsymbol{\nabla} s$, over an entire flux surface vanishes \cite{helander2014theory}. Thus, in the absence of collisions and turbulence, passing particles are confined. To study fast-ion confinement, the focus should therefore be on trapped particles, whose net radial displacement is non-zero in general. In the presence of collisions, passing particles can become trapped and get lost, but at such a slow rate that it is not relevant for the discussion of this section.

To describe the dynamics of trapped particles at time scales longer than the time that it takes them to complete a bounce orbit, the second adiabatic invariant
\begin{equation}
    J \! \left( s , \alpha , \mathcal{E} , \mu \right) = 2 \! \int_{l_{b_1}}^{l_{b_2}} \! \left| v_{\|} \right| \hspace{0.03 cm} \mathrm{d} l
\end{equation}
is usually employed. Here, the integral over the arc-length is taken between the bounce points $l_{b_1}$ and $l_{b_2}$. The second adiabatic invariant is related to the drift velocity through the identities
\begin{equation} \label{eq_relation_d_a_J}
    \partial_{\alpha} J = \frac{Z e \Psi_{LCFS}}{m} \tau_b \overline{\mathbf{v}_d \cdot \boldsymbol{\nabla} s} ,
\end{equation}
\begin{equation} \label{eq_relation_d_s_J}
    \partial_s J = - \frac{Z e \Psi_{LCFS}}{m} \tau_b \overline{\mathbf{v}_d \cdot \boldsymbol{\nabla} \alpha} ,
\end{equation}
where $\overline{(\cdot)}$ denotes bounce-average, which, for functions $f \! \left( s , \alpha , l , \mathcal{E} , \mu \right)$ that do not depend on $\sigma$, is expressed as
\begin{equation}
    \overline{f} = \frac{2}{\tau_b} \int_{l_{b_1}}^{l_{b_2}} \!\! f \hspace{0.06 cm} \frac{\mathrm{d} l}{\left| v_{\|} \right|} ,
\end{equation}
where
\begin{equation}
    \tau_b = 2 \! \int_{l_{b_1}}^{l_{b_2}} \!\! \hspace{0.06 cm} \frac{\mathrm{d} l}{\left| v_{\|} \right|}
\end{equation}
is the bounce time.

Equations (\ref{eq_relation_d_a_J}) and (\ref{eq_relation_d_s_J}) imply that trapped particles move at constant $J$ (in the absence of collisions and turbulence). It automatically follows that, if constant-$J$ contours are aligned with flux surfaces, trapped particles are well confined because, then, $\partial_{\alpha} J = 0$ and $\overline{\mathbf{v}_d \cdot \boldsymbol{\nabla} s} = 0$. More importantly in practice (since deviations from omnigeneity are inevitable), good trapped-particle confinement could also be achieved if constant-$J$ contours are closed and do not intersect the LCFS. This is the case when, even if $\partial_{\alpha} J$ is not zero, $\left| \partial_{\alpha} J \right| \ll \left| \partial_s J \right|$, which, according to equations (\ref{eq_relation_d_a_J}) and (\ref{eq_relation_d_s_J}), is equivalent to
\begin{equation} \label{eq_good_confinement}
    \left| \overline{\mathbf{v}_d \cdot \boldsymbol{\nabla} s} \right| \ll \left| \overline{\mathbf{v}_d \cdot \boldsymbol{\nabla} \alpha} \right| .
\end{equation}

In other words, even when $\overline{\mathbf{v}_d \cdot \boldsymbol{\nabla} s} = 0$ is not satisfied, good fast-ion confinement could be accomplished by achieving a large $\left| \overline{\mathbf{v}_d \cdot \boldsymbol{\nabla} \alpha} \right|$. Indeed, accordingly, some previous works \cite{velasco2021model,nemov2008poloidal} employ the ratio $\left| \overline{\mathbf{v}_d \cdot \boldsymbol{\nabla} s} \right| \! / \! \left| \overline{\mathbf{v}_d \cdot \boldsymbol{\nabla} \alpha} \right|$ to define figures of merit for fast-ion confinement in the framework of stellarator optimization.

In light of this, let us give more explicit expressions for the bounce-average of the tangential component of the magnetic and $E \times B$ drifts. Assuming that the electrostatic potential is a flux-function, $\Phi = \Phi \! \left( s \right)$, the electric field is radial, $\mathbf{E} = E_s \! \left( s \right) \boldsymbol{\nabla} s$. Taking the dot product of equation (\ref{eq_def_vE}) and $\boldsymbol{\nabla} \alpha$ one obtains ${\mathbf{v}_E \cdot \boldsymbol{\nabla} \alpha} = - E_s \! \left( s \right) / \Psi_{LCFS}$ and, therefore,
\begin{equation} \label{eq_vE_alpha_avrg}
    \overline{\mathbf{v}_E \cdot \boldsymbol{\nabla} \alpha} = I_{E_s} E_s \! \left( s \right) ,
\end{equation}
where
\begin{equation}
    I_{E_s} = - \frac{1}{\Psi_{LCFS}} .
\end{equation}
Hence, the bounce-average of the tangential component of the $E \times B$ drift depends only on the value of $E_s$ on the surface in question. When taking the dot product of equation (\ref{eq_def_vM}) and $\boldsymbol{\nabla} \alpha$ and performing the bounce-average, the result is
\begin{equation} \label{eq_vM_alpha_avrg}
    \overline{\mathbf{v}_M \cdot \boldsymbol{\nabla} \alpha} = \overline{\frac{2 \mathcal{E}}{Z e \Psi_{LCFS}} \left( 1 - \frac{\lambda B}{2} \right) \frac{\partial_s B}{B}} .
\end{equation}

The term $\partial_s B / B$ in equation (\ref{eq_vM_alpha_avrg}) depends on $\beta = 2 \mu_0 p / B^2$, where $p$ is the plasma pressure and $\mu_0$ is the vacuum magnetic permeability. In large aspect ratio stellarators, the effect of increasing $\beta$ can be approximately absorbed in the derivative
\begin{equation}
    \frac{\partial_s B}{B} \approx \frac{\partial_s B_{vac}}{B_{vac}} - \frac{\partial_s \beta}{2} ,
\end{equation}
where $B_{vac}$ is the vacuum magnetic field strength \cite{herbemont2022finite}. The bounce-average of the tangential magnetic drift can thus be divided into two contributions,
\begin{equation} \label{eq_vM_alpha_avrg_def}
    \overline{\mathbf{v}_M \cdot \boldsymbol{\nabla} \alpha} = I_0 + I_{\partial_s \beta} \partial_s \beta ,
\end{equation}
where
\begin{equation}
    I_0 = \frac{4 \mathcal{E}}{Z e \Psi_{LCFS} \tau_b} \int_{l_{b_1}}^{l_{b_2}} \!\! \hspace{0.06 cm} \left( 1 - \frac{\lambda B_{vac}}{2} \right) \frac{\partial_s B_{vac}}{B_{vac}} \frac{\mathrm{d} l}{v_{\|}}
\end{equation}
and
\begin{equation}
    I_{\partial_s \beta} = - \frac{2 \mathcal{E}}{Z e \Psi_{LCFS} \tau_b} \int_{l_{b_1}}^{l_{b_2}} \!\! \hspace{0.06 cm} \left( 1 - \frac{\lambda B_{vac}}{2} \right) \frac{\mathrm{d} l}{v_{\|}} ,
\end{equation}
and all the magnitudes in the integrals are evaluated with the vacuum magnetic field.

Combining equations (\ref{eq_vE_alpha_avrg}) and (\ref{eq_vM_alpha_avrg_def}), it follows that the bounce-average of the total tangential drift is
\begin{equation} \label{eq_vd_alpha_avrg}
    \overline{\mathbf{v}_d \cdot \boldsymbol{\nabla} \alpha} = I_0 + I_{\partial_s \beta} \partial_s \beta + I_{E_s} E_s .
\end{equation}

In section \ref{academic_scan}, linear profiles (in $s$) for $\beta \! \left( s \right)$ are assumed, $\beta \! \left( s \right) = 2 \! \left \langle \beta \right \rangle \! \left( 1 - s \right)$, so $\partial_s \beta = - 2 \! \left \langle \beta \right \rangle$. Therefore, the bounce-average of the total tangential drift can be expressed as
\begin{equation} \label{eq_vd_alpha_avrg_def}
    \overline{\mathbf{v}_d \cdot \boldsymbol{\nabla} \alpha} = I_0 + I_{\left \langle \beta \right \rangle} \! \left \langle \beta \right \rangle + I_{E_s} E_s ,
\end{equation}
where
\begin{equation}
    I_{\left \langle \beta \right \rangle} = \frac{4 \mathcal{E}}{Z e \Psi_{LCFS} \tau_b} \int_{l_{b_1}}^{l_{b_2}} \!\! \hspace{0.06 cm} \left( 1 - \frac{\lambda B_{vac}}{2} \right) \frac{\mathrm{d} l}{v_{\|}}
\end{equation}
and, again, all the magnitudes in the integral are evaluated with the vacuum magnetic field. The integrals $I_0$, $I_{\partial_s \beta}$ and $I_{\left \langle \beta \right \rangle}$ depend spatially on $s$ and $\alpha$. Whereas $I_0$ can be positive or negative, since the integrand (specifically the quotient $\partial_s B_{vac} / B_{vac}$) can be positive or negative, $I_{\left \langle \beta \right \rangle}$ can only be positive, since the integrand is always positive. Moreover, in W7-X or in most QI stellarators, $\partial_s B_{vac} / B_{vac}$ is relatively small, thus, $I_0$ is small compared to $I_{\left \langle \beta \right \rangle} \! \left \langle \beta \right \rangle$ for finite $\left \langle \beta \right \rangle$ (or to $I_{\partial_s \beta} \partial_s \beta$ for finite $\partial_s \beta$) and $\overline{\mathbf{v}_M \cdot \boldsymbol{\nabla} \alpha}$ can only be large through a finite $\left \langle \beta \right \rangle$ (or a finite $\partial_s \beta$).

Equation (\ref{eq_vd_alpha_avrg_def}) shows that the bounce-average of the tangential drift depends linearly on $\left \langle \beta \right \rangle$ and the radial electric field. Therefore, with respect to $\overline{\mathbf{v}_d \cdot \boldsymbol{\nabla} \alpha}$, a large $I_{E_s} E_s$ is equivalent to a large $I_{\left \langle \beta \right \rangle} \! \left \langle \beta \right \rangle$ and both can lead to condition (\ref{eq_good_confinement}). Hence, increasing the bounce-average of the tangential drift either by varying $\left \langle \beta \right \rangle$ or the radial electric field should reduce fast-ion losses.

The balance among the different terms depends on the type of machine and, especially, its size. The size of the radial electric field typically scales with the ion temperature gradient, and this determines the magnitude of $I_{E_s} E_s$, while the size of $I_0$ and $I_{\left \langle \beta \right \rangle}$ increases with the fast-ion energy. In a reactor, $I_0$ and $I_{\left \langle \beta \right \rangle}$ dominate the bounce-average of the tangential drift. Specifically, in a QI reactor, $I_0$ is small and, therefore, $I_{\left \langle \beta \right \rangle} \! \left \langle \beta \right \rangle$ is the dominant term. Thereby a high $\left \langle \beta \right \rangle$ is required for a good fast-ion confinement. On the other hand, in W7-X, since the ratio of the fast-ion energy to that of the bulk ions is much lower than in a reactor, $I_{E_s} E_s$ is relevant and tends to dominate $\overline{\mathbf{v}_d \cdot \boldsymbol{\nabla} \alpha}$. Even at high $\left \langle \beta \right \rangle$, $I_{E_s} E_s$ is comparable to $I_{\left \langle \beta \right \rangle} \! \left \langle \beta \right \rangle$.

The ideal scan on $\left \langle \beta \right \rangle$ mentioned in the introduction is thus specifically a scan on the size of $I_{\left \langle \beta \right \rangle} \! \left \langle \beta \right \rangle$ at negligible $I_{E_s} E_s$. However, such a scan, specially the aspect of small $I_{E_s} E_s$, is difficult to perform in the experiment. What is proposed instead in this work is a controlled scan on $I_{E_s} E_s$ (or, even if $\left \langle \beta \right \rangle$ cannot be maintained constant, a scan on $I_{\left \langle \beta \right \rangle} \! \left \langle \beta \right \rangle + I_{E_s} E_s$) for the validation of the enhanced fast-ion confinement at large $\overline{\mathbf{v}_d \cdot \boldsymbol{\nabla} \alpha}$.

\section{Methods and Workflow}
\label{workflow}

In this work, a numerical approach is employed. The fast-ion loss fraction is computed via collisional guiding-center simulations of fast-ion trajectories, using the ASCOT5 orbit-following code \cite{varje2019high,varje2022energetic,sarkimaki2019modelling,hirvijoki2014ascot}. All simulations are based on the KJM (high mirror) configuration of W7-X. For each case, the trajectories of $10^4$ particles are simulated. Simulated fast ions are hydrogen nuclei ($^1$H$^+$) with an initial energy of $50 \ \mathrm{keV}$, in the range that is accessible with the heating systems of W7-X. They present a normalized Larmor radius similar to that of alpha particles in a reactor.

From the simulation outcome, only the initial and final states of the particles are investigated, to determine the different results addressed: (energy) loss fraction and (energy) loss fraction distribution in pitch-angle. The energy loss fraction is defined as the sum of the energy of particles that scape the plasma divided by the sum of the original energy of all simulated particles.

For each of the scenarios considered in this work, we use three different initial distributions, which are isotropic in velocity, with a fixed initial energy of $50 \ \mathrm{keV}$ and with initial positions uniformly distributed on one flux surface. The difference between the three distributions lies in the surface where the particles are born. We choose the initial surfaces to be $\rho_0 = \left \{ 0.25, 0.50, 0.71 \right \}$, where $\rho \equiv \sqrt{s}$ is the radial variable that we will use from now on.

Particles are simulated until they escape or thermalize. In the simulation, particles are considered to have escaped from the plasma when they cross the LCFS. Similarly, a particle is considered to have thermalized if its energy is reduced below $1 \ \mathrm{keV}$ or below twice the bulk-ion temperature.

The magnetic configuration is computed via free-boundary simulations with the VMEC code \cite{hirshman1983steepest,hirshman1985optimized,lazerson2016verification}. This code takes as input the pressure profile, which is obtained from the density and temperature profiles, and the information about the coil currents. In this work, only one ion species in the bulk plasma is considered: the hydrogen nucleus ($^1$H$^+$). Therefore, the expression for the pressure reads $p = n_{e} T_{e} + n_{i} T_{i}$. Here, $n_e$ is the electron density, $T_e$ is the electron temperature, $n_i$ is the ion density and $T_i$ is the ion temperature. Quasi-neutrality is assumed, thus, the ion density is directly related to the electron density via $n_{e} = n_{i}$ (fast ions are considered trace species). Once the calculation of the equilibrium via VMEC is completed, it is exported to the ASCOT5 suite of codes using the EXTENDER code \cite{drevlak2005pies}. The same density and temperature profiles are also used to calculate the collisions.

Two types of scans will be performed in the next sections. In the academic scan on $\left \langle \beta \right \rangle$ and $E_s$ that we will discuss in section \ref{academic_scan} the radial electric field profile is imposed without asking for consistency with the plasma profiles (i.e., without requiring ambipolarity of the neoclassical fluxes). In the experimentally-based scan addressed in section \ref{experiment_scan}, the radial electric field profile is computed via the SFINCS code \cite{landreman2014comparison,landreman2013new}, using the density and temperature profiles and the magnetic configuration, previously calculated, as input.

\begin{figure*}[ht!]
    \centering
    \includegraphics[width=0.95\linewidth]{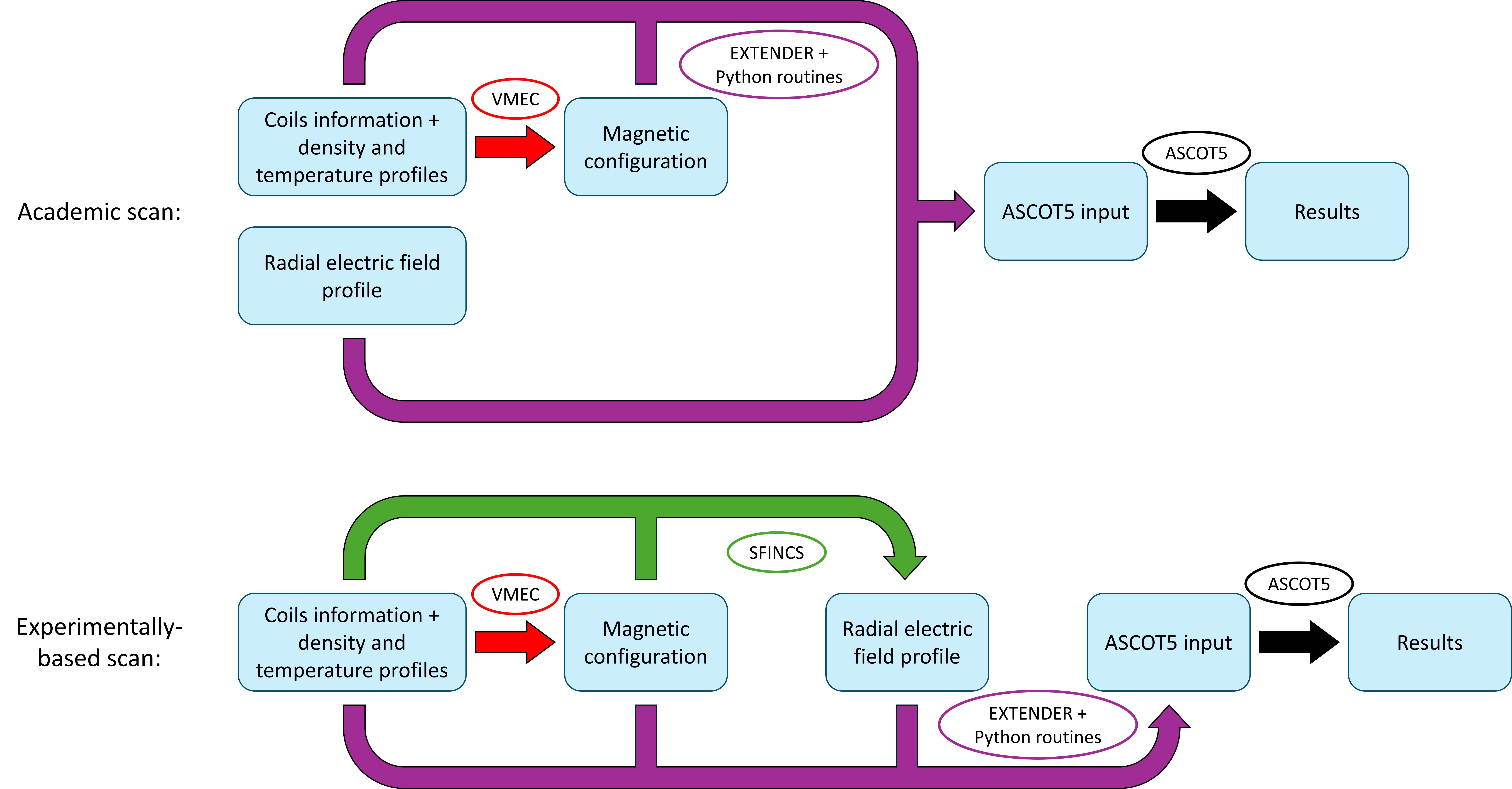}
    \hspace{0.3 cm}
    \caption{Diagram of the workflow followed in the simulations scans from this work, both for the academic scan (top) and the experimentally-based scan (bottom).}
    \label{Fig_3_01_workflow_diag}
\end{figure*}

In order to facilitate the understanding of the workflow, a diagram for both the academic scan discussed in section \ref{academic_scan} and the experimentally-based scan addressed in section \ref{experiment_scan} is provided in figure \ref{Fig_3_01_workflow_diag}.

\section{Academic scan}
\label{academic_scan}

In this section, we aim to prove that, regarding fast-ion confinement, an increase in $\left \langle \beta \right \rangle$ produces a strictly equivalent effect to an increase in the absolute value of the radial electric field. In this first scan, we choose unrealistic plasma profiles in order to illustrate the aforementioned equivalence more clearly. We will turn to realistic profiles in section \ref{experiment_scan}.

\begin{figure*}[ht!]
    \centering
    \includegraphics[width=0.32\linewidth]{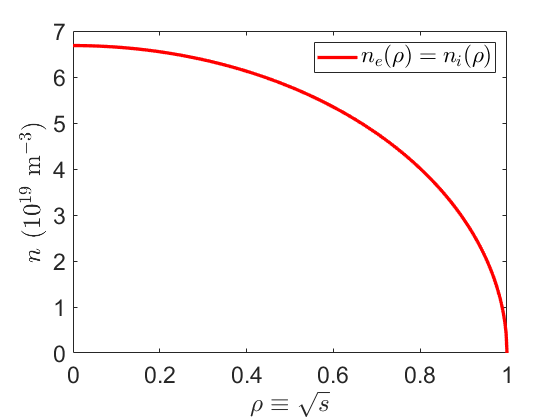}
    \hspace{0.05 cm}
    \includegraphics[width=0.32\linewidth]{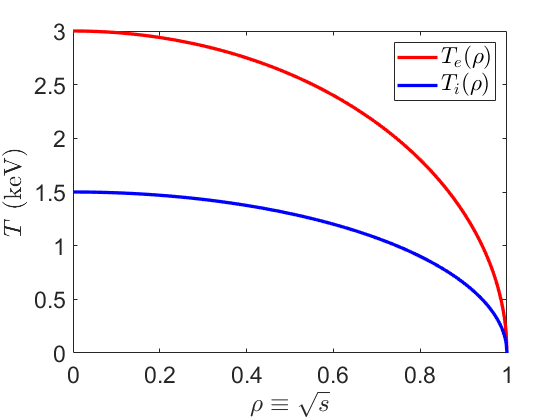}
    \hspace{0.05 cm}
    \includegraphics[width=0.32\linewidth]{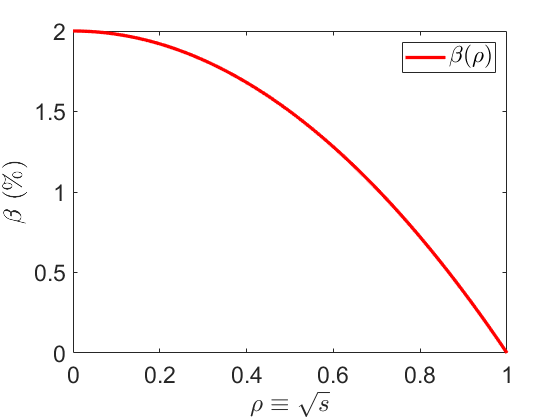}
    \hspace{0.05 cm}
    \caption{Density (left), temperature (center) and $\beta$ (right) profiles for the case $\left \langle \beta \right \rangle = 1 \%$.}
    \label{Fig_4_01_n_T_b_profiles}
\end{figure*}

The density and temperature profiles are proportional to $\sqrt{1 - \rho^2}$,
\begin{equation}
\begin{gathered}
    n_e = n_{e0} \sqrt{1 - \rho^2} , \\
    n_i = n_{i0} \sqrt{1 - \rho^2} , \\
    T_e = T_{e0} \sqrt{1 - \rho^2} , \\
    T_i = T_{i0} \sqrt{1 - \rho^2} ,
\end{gathered}
\end{equation}
so that $\beta$ profiles are parabolic in $\rho$ (linear in $s$), as shown in figure \ref{Fig_4_01_n_T_b_profiles}. Therefore, the derivative $\partial_s \beta$, which plays an important role in $\overline{\mathbf{v}_M \cdot \boldsymbol{\nabla} \alpha}$, is radially-constant and proportional to $\left \langle \beta \right \rangle$. Because of quasi-neutrality, it follows that $n_{i0} = n_{e0}$. For simplicity, the temperature profiles are kept fixed during the scan, with $T_{e0} = 3.0 \ \mathrm{keV}$ and $T_{i0} = 1.5 \ \mathrm{keV}$. Thus, there is only one free parameter, $n_{e0}$. By assigning different values to $n_{e0}$, different values of $\left \langle \beta \right \rangle$ are obtained.

\begin{figure*}[ht!]
    \centering
    \includegraphics[width=0.37\linewidth]{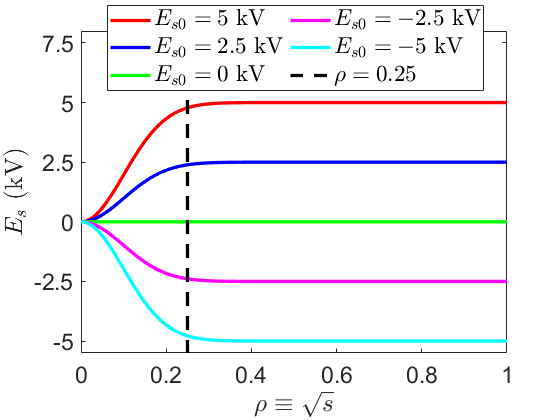}
    \hspace{0.3 cm}
    \includegraphics[width=0.37\linewidth]{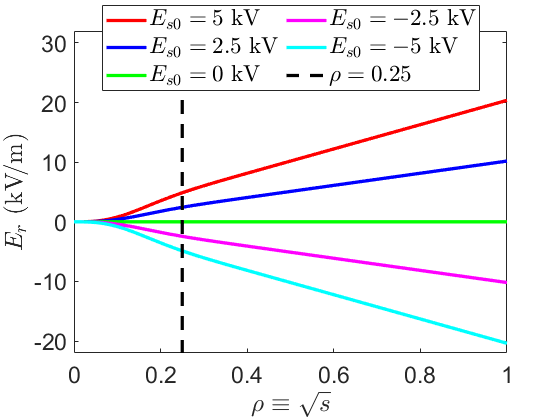}
    \hspace{0.15 cm}
    \caption{$E_s$ (left) and $E_r$ (right) profiles for several cases in the academic scan.}
    \label{Fig_4_02_Es_Er_profiles}
\end{figure*}

For the radial electric field, almost flat $E_s$ profiles are chosen because they produce flat profiles of $\overline{\mathbf{v}_E \cdot \boldsymbol{\nabla} \alpha}$. The analytical expression for these profiles is
\begin{equation}
    E_s = E_{s0} \left( 1 - e^{- s / C} \right) ,
\end{equation}
with $C = 0.02$, ensuring the profile is flat from $s = 0.06$ (i.e., $\rho = 0.25$) outwards. The relation between $E_r$ and $E_s$ is $E_r = \left( 2 \rho / a \right) E_s$, where $a$ is the minor radius, which is $0.49 \ \mathrm{m}$ for the high mirror configuration. Hence, flat $E_s$ profiles imply linear $E_r$ profiles. In figure \ref{Fig_4_02_Es_Er_profiles}, both the $E_s$ and $E_r$ profiles used in this section are represented for several values of $E_{s0}$.

This choice of the profiles ensures that $\partial_s \beta$ and $E_s$ do not depend on the surface, thus the relative size of the different terms in equation (\ref{eq_vd_alpha_avrg_def}) does not change significantly across different surfaces. This could complicate the analysis, as we will discuss in section \ref{experiment_scan}.

A scan on $\left \langle \beta \right \rangle$ from $\left \langle \beta \right \rangle = 0.01 \%$ to $\left \langle \beta \right \rangle = 4 \%$ and on $E_{s0}$ from $E_{s0} = - 5 \ \mathrm{kV}$ to $E_{s0} = 5 \ \mathrm{kV}$ is carried out, performing the simulation of fast-ion trajectories for each scenario (each $\left \langle \beta \right \rangle$ and $E_{s0}$).

\begin{figure*}[ht!]
    \centering
    \includegraphics[width=0.322\linewidth]{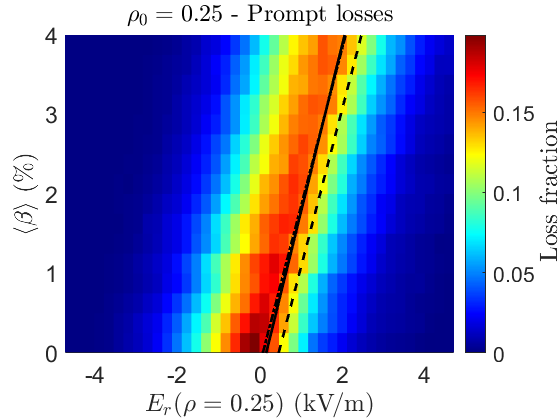}
    \hspace{0.08 cm}
    \includegraphics[width=0.322\linewidth]{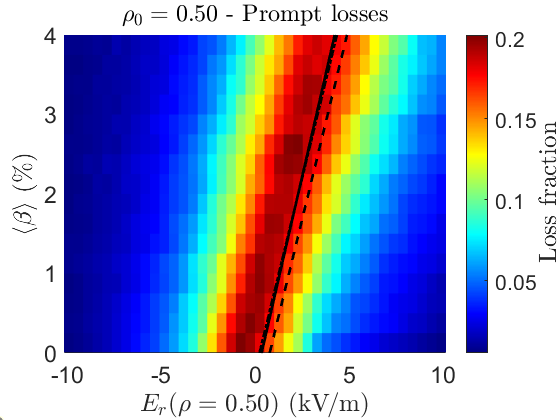}
    \hspace{0.08 cm}
    \includegraphics[width=0.322\linewidth]{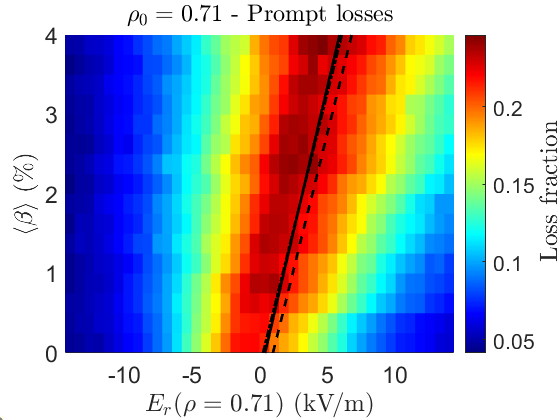}
    \caption{Energy prompt loss fraction as a function of $\left \langle \beta \right \rangle$ and $E_r$ for $\rho_0 = 0.25$ (left), $\rho_0 = 0.50$ (center) and $\rho_0 = 0.71$ (right). The region of $\overline{\mathbf{v}_d \cdot \boldsymbol{\nabla} \alpha} \approx 0$, computed with the theoretical model explained above, is represented with lines for moderately-trapped (solid line), barely-trapped (dashed line) and deeply-trapped (dash-dotted line) particles.}
    \label{Fig_4_03_f_prompt_vs_er_beta}
\end{figure*}

Figure \ref{Fig_4_03_f_prompt_vs_er_beta} shows the prompt loss fraction as a function of $\left \langle \beta \right \rangle$ and $E_r$. Prompt losses are defined here as the losses occurring in $10^{-3} \ \mathrm{s}$ after the birth of the fast ions. Figure \ref{Fig_4_03_f_prompt_vs_er_beta} shows a linear relationship between $\left \langle \beta \right \rangle$ and the value of $E_r$ that produces maximal losses. This is consistent with the hypothesis that the maximum in losses for a given magnetic configuration is connected to a slow poloidal precession of the fast ions, and with the prediction of equation (\ref{eq_vd_alpha_avrg_def}) that $\overline{\mathbf{v}_d \cdot \boldsymbol{\nabla} \alpha} \approx 0$ for a straight line in the $\left( E_r , \! \left \langle \beta \right \rangle \right)$ parameter space. Figure \ref{Fig_4_03_f_prompt_vs_er_beta} also shows that there is a monotonous reduction in losses as $E_r$ departs from the value that produces maximal losses, to both the positive and negative signs. This can be understood as this departure of $E_r$ increases $\left| \overline{\mathbf{v}_d \cdot \boldsymbol{\nabla} \alpha} \right|$, approaching condition (\ref{eq_good_confinement}). Equation (\ref{eq_vd_alpha_avrg_def}) predicts that ${\overline{\mathbf{v}_d \cdot \boldsymbol{\nabla} \alpha} \approx 0}$ for $E_r = \frac{2 \rho}{a} \Psi_{LCFS} \left( I_0 + I_{\left \langle \beta \right \rangle} \left \langle \beta \right \rangle \right)$. The fact that $I_0 + I_{\left \langle \beta \right \rangle} \left \langle \beta \right \rangle$ is generally non-zero explains, through equation (\ref{eq_vd_alpha_avrg_def}), why the effects of radial electric fields of equal magnitude but opposite sign are not the same, see e.g. \cite{green2026energetic}.

We now approximate the field of W7-X to an exactly QI field. This is done in order to remove the $\alpha$ dependence of $I_0$ and $I_{\left \langle \beta \right \rangle}$, which are flux-surface constants in a QI field. We then compute the integrals $I_0$ and $I_{\left \langle \beta \right \rangle}$ analytically, as shown in (\ref{eq_I_0_model}) and (\ref{eq_I_b_model}). In effect, as can be seen in figure \ref{Fig_4_03_f_prompt_vs_er_beta}, simulations with ASCOT exhibit maximal losses in a region of the parameter space consistent with where the theoretical model predicts $\overline{\mathbf{v}_d \cdot \boldsymbol{\nabla} \alpha} \approx 0$. There is a good agreement (in sign and size) on the slope of the region of maximal losses. However, there is a small discrepancy (in absolute terms) on the y-intercept (the value of $E_r$ that produces maximal losses for $\left \langle \beta \right \rangle = 0$).

\begin{figure*}[ht!]
    \centering
    \includegraphics[width=0.825\linewidth]{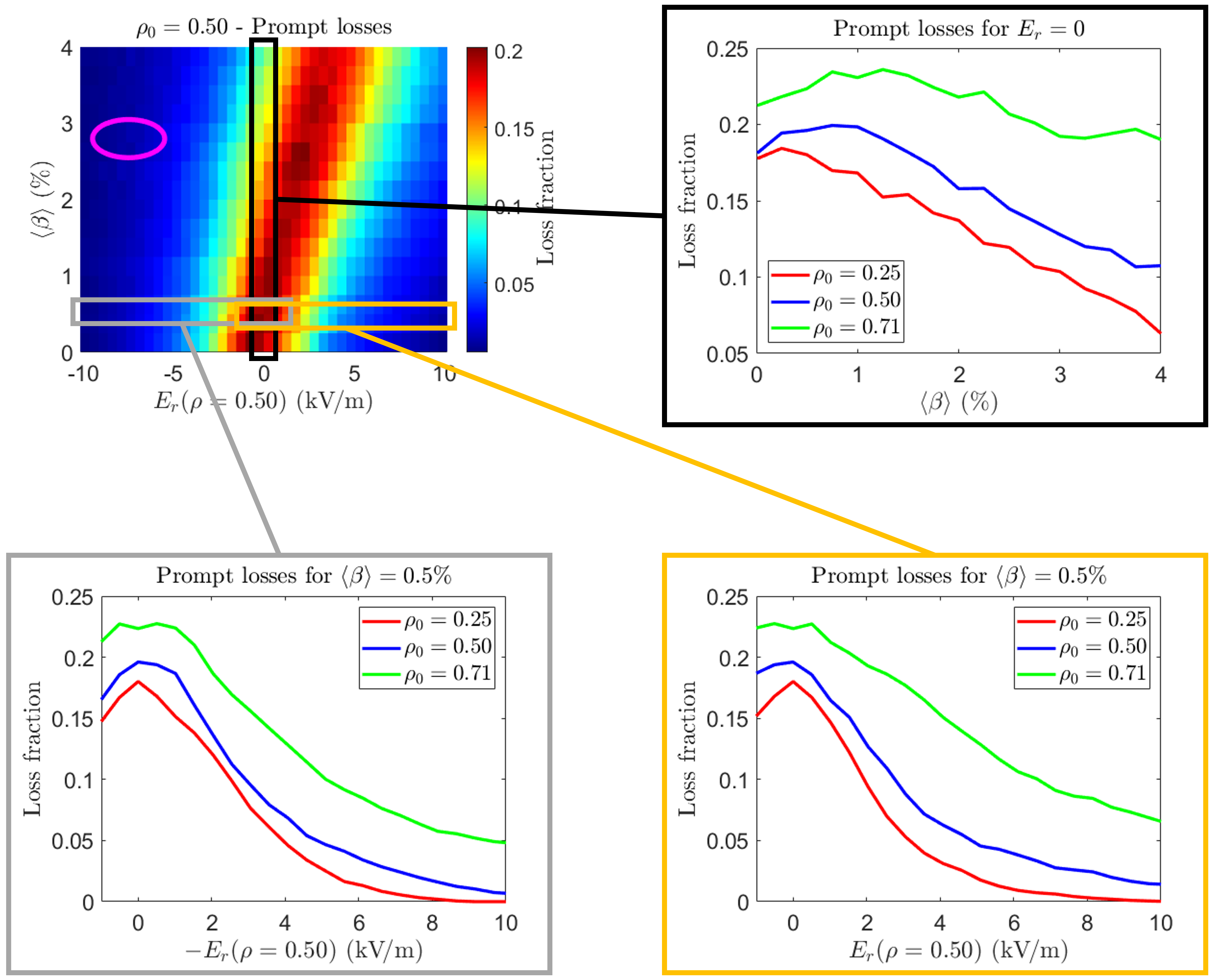}
    \caption{Energy prompt loss fraction as a function of $\left \langle \beta \right \rangle$ and $E_r$ for $\rho_0 = 0.50$ with several regions highlighted (top left), energy prompt loss fraction as a function of $\left \langle \beta \right \rangle$ in the absence of electric field (top right), energy prompt loss fraction as a function of $E_r$ for $\left \langle \beta \right \rangle = 0.5 \%$ for negative values of $E_r$ (bottom left) and for positive values of $E_r$ (bottom right). Values of $\left \langle \beta \right \rangle$ and $E_r$ for high-$\beta$ discharges are shown with a magenta ellipse in the top left plot.}
    \label{Fig_4_04_f_prompt_vs_er_or_beta}
\end{figure*}

From the previous 2D scan, two 1D scans can be extracted, as shown in figure \ref{Fig_4_04_f_prompt_vs_er_or_beta}. First, a scan on $\left \langle \beta \right \rangle$ in the absence of electric field, which is a vertical cut in the plots from figure \ref{Fig_4_03_f_prompt_vs_er_beta}. Second, a scan on the radial electric field with a constant value of $\left \langle \beta \right \rangle = 0.5 \%$, which is a horizontal cut in the plots from figure \ref{Fig_4_03_f_prompt_vs_er_beta}.

As found in previous works \cite{drevlak2014fast,velasco2021model}, in the absence of radial electric field, fast-ion losses decrease when $\left \langle \beta \right \rangle$ is increased enough (from $\left \langle \beta \right \rangle \sim 0.5 \%$ on). On the other hand, when the effect of the radial electric field is considered at constant $\left \langle \beta \right \rangle$, one can see that there is a value of $E_r$ that produces maximal losses and a reduction in losses when $E_r$ departs from this value. If the ranges are chosen properly, as in figure \ref{Fig_4_04_f_prompt_vs_er_or_beta}, the dependence of losses on $\left \langle \beta \right \rangle$ is similar to their dependence on $E_r$.

Hence, these results support the fact that the effects of $\left \langle \beta \right \rangle$ and $E_r$ on $\overline{\mathbf{v}_d \cdot \boldsymbol{\nabla} \alpha}$ are quantitatively equivalent, as depicted in equation (\ref{eq_vd_alpha_avrg_def}), and that the condition (\ref{eq_good_confinement}) is crucial in fast-ion confinement.

With the results shown in this section, we see that, regarding fast-ion confinement in the high mirror configuration, the ideal scan on $\left \langle \beta \right \rangle$ could be replaced with an ideal scan on the radial electric field. In the next section, we investigate whether it is possible to perform such a scan (or a sufficiently similar one) in the experiment.

Finally, discharges with relatively high values of $\left \langle \beta \right \rangle$ have been recently obtained in the experiment \cite{grulke2025overview}. These values of $\left \langle \beta \right \rangle$ and the typical values of the radial electric field in those discharges are shown in figure \ref{Fig_4_04_f_prompt_vs_er_or_beta} (with a magenta ellipse). However, we note that trying to use this type of discharges to experimentally study fast-ion confinement could be problematic because of two reasons. First, there is a large improvement in fast-ion confinement compared to the case of $\left \langle \beta \right \rangle = 0$ and $E_r = 0$. However, this improvement is arguably mainly due to the finite radial electric field and not to the finite $\left \langle \beta \right \rangle$. A purely “vertical” scan, in which the effect of $E_r$ is, if not negligible, at least constant, is complicated. Second, those discharges are located in a region where losses are significantly small, which may e.g. lead to a too low signal in fast-ion loss detectors. In the fine scan on the radial electric field at constant $\left \langle \beta \right \rangle$ that we are proposing, studying parametric dependencies may be, in principle, easier.

\section{Experimentally-based scan}
\label{experiment_scan}

In this section, experimentally-based profiles are employed to perform a scan on the radial electric field. As this quantity cannot be directly controlled in the experiment, we use different density and temperature profiles to generate, through the ambipolarity condition, different radial electric field profiles. These density and temperature profiles come from 37 different time instants of the experimental discharge \#20181009.034, which is a turbulence-reduced discharge with increased confinement \cite{ford2024turbulence}. The plasma starts, as is usually the case, with gas fuelling and $2 \ \mathrm{MW}$ of X2-mode Electron Cyclotron Resonance Heating (ECRH), achieving a typical plasma with high electron and limited ion temperature at moderate density, before switching to $2.5 \ \mathrm{MW}$ pure NBI heating. At the transition, the electron temperature decreases until it equals that of the ions. Then, the density, whose profile becomes more peaked, continues to increase and both temperatures rise slightly, up to saturation at $1.3 \ \mathrm{keV}$. Just before the end of the pure NBI phase, the density reaches a maximum of $12 \times 10^{19} \ \mathrm{m}^{-3}$ and its profile becomes strongly peaked. With the reintroduction of an additional $1 \ \mathrm{MW}$ of O2-mode ECRH, both temperatures increase rapidly (the ion temperature peaks at $1.75 \ \mathrm{keV}$, whereas the electron temperature continues to increase to $1.85 \ \mathrm{keV}$), while the density decreases over $500 \ \mathrm{ms}$, reaching $10 \times 10^{19} \ \mathrm{m}^{-3}$, which is still a relatively high density. As the density decreases, the density gradient also relaxes due to the effect of ECRH, leading to the reappearance of turbulence and a reduction in confinement. After this $500 \ \mathrm{ms}$ interval, the temperatures also relax to around the typical value of $1.6 \ \mathrm{keV}$, until the end of the discharge.

\begin{figure*}[ht!]
    \centering
    \includegraphics[width=0.4\linewidth]{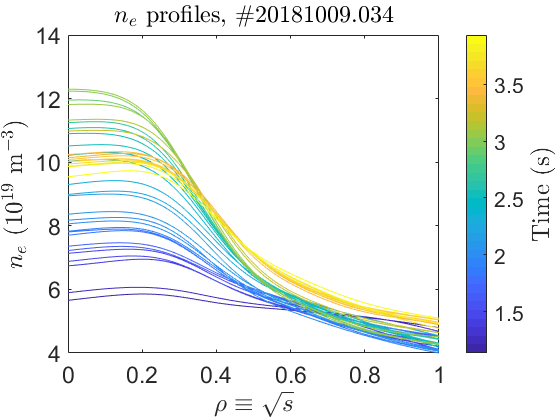}
    \hspace{0.5 cm}
    \vspace{0.4 cm}
    \includegraphics[width=0.4\linewidth]{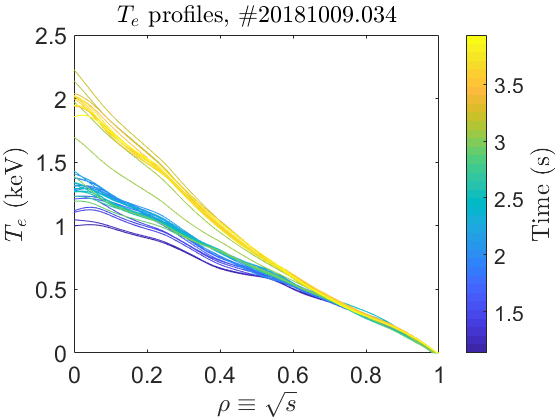}
    \includegraphics[width=0.4\linewidth]{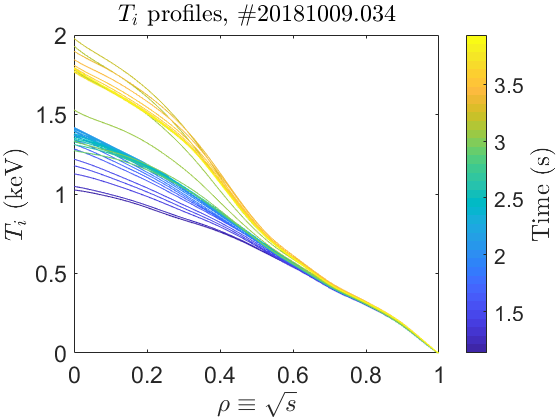}
    \hspace{0.5 cm}
    \includegraphics[width=0.4\linewidth]{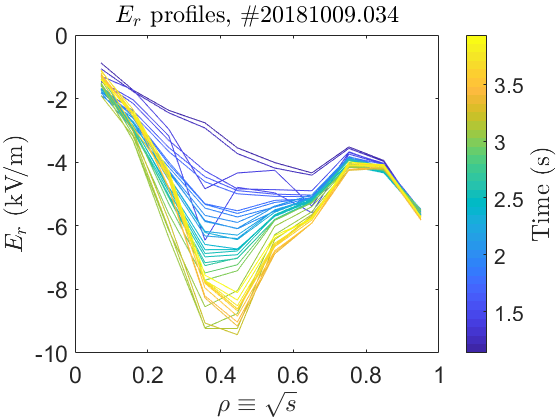}
    \caption{Electron density (top left), electron temperature (top right), ion temperature (bottom left) and radial electric field (bottom right) profiles from the 37 different time instants of the experimental discharge \#20181009.034.}
    \label{Fig_5_01_n_T_Er_profiles}
\end{figure*}

The profiles from this discharge are displayed in figure \ref{Fig_5_01_n_T_Er_profiles}. In particular, figure \ref{Fig_5_01_n_T_Er_profiles} (bottom right) shows the radial electric field profiles computed with the SFINCS code, as explained in section \ref{workflow}. All of them are negative throughout the entire range of $\rho$ and are similar in shape, which simplifies the analysis of their effect on fast-ion confinement. In this section, the value of $E_r$ at $\rho = 0.45$ is used as the quantity that represents the radial electric field profile size.

\begin{figure*}[ht!]
    \centering
    \includegraphics[width=0.875\linewidth]{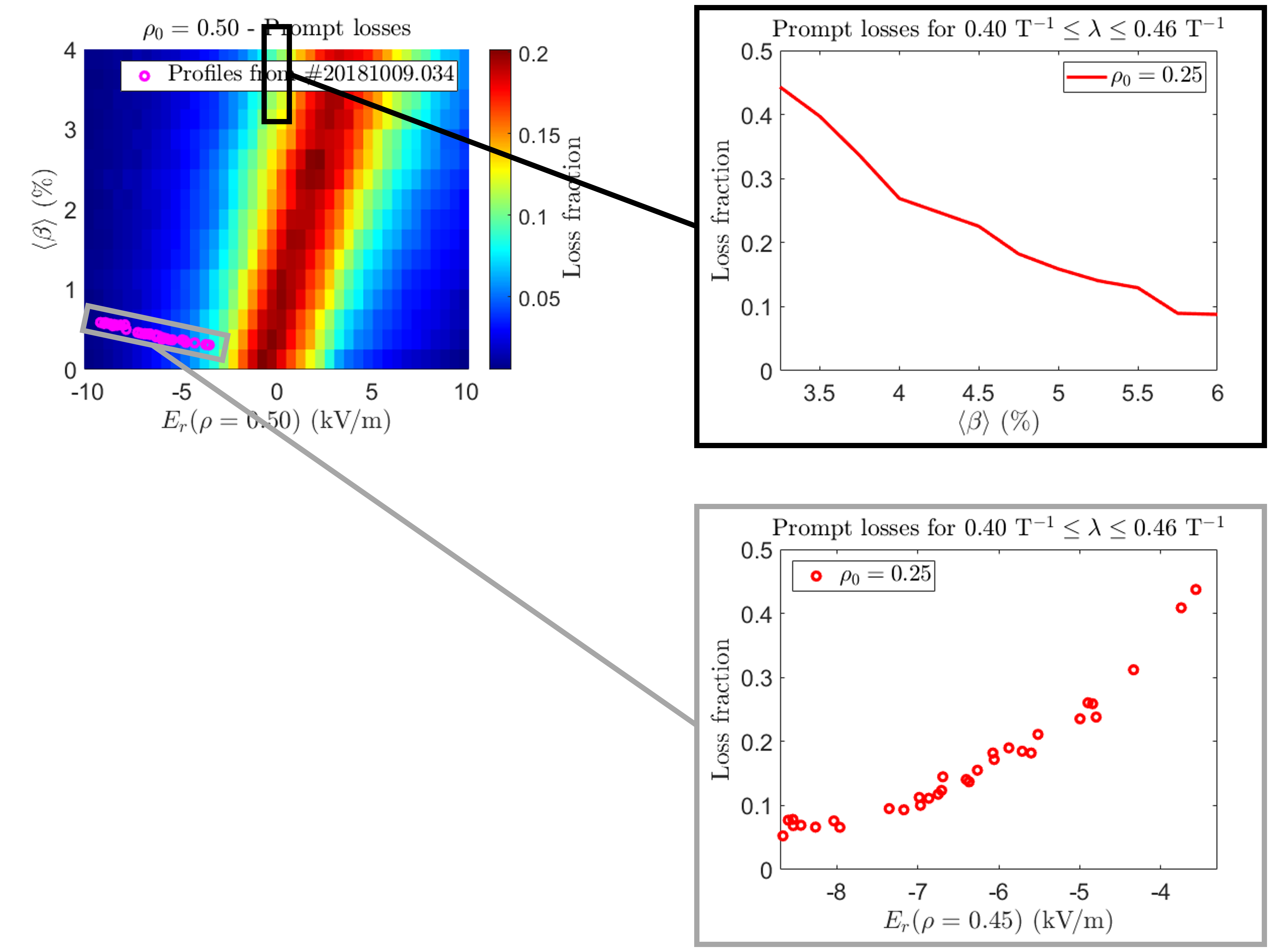}
    \hspace{0.4 cm}
    \caption{Region of the academic scan explored by the selected plasmas (left). Energy prompt loss fraction as a function of $\left \langle \beta \right \rangle$ from the ideal scan (top right) and as a function of $E_r$ at $\rho = 0.45$ from the experimental profiles (bottom right). Both of them are computed for for deeply-trapped particles with $0.40 \ \mathrm{T}^{-1} \leq \lambda \leq 0.46 \ \mathrm{T}^{-1}$ and $\rho_0 = 0.25$.}
    \label{Fig_5_04_losses_deeply_trapped_comp}
\end{figure*}

After computing the value of $\left \langle \beta \right \rangle$ for each case (and neglecting the fact that the profiles are not parabolic), in figure \ref{Fig_5_04_losses_deeply_trapped_comp} (left) we can see that the plasmas of \#20181009.034 are located at the bottom left part of the plots from figure \ref{Fig_4_03_f_prompt_vs_er_beta}, relatively close to the maximal losses region. It can also be noticed that the change in $\left \langle \beta \right \rangle$ throughout the scan is small ($\left \langle \beta \right \rangle _{max} - \left \langle \beta \right \rangle _{min} \simeq 0.25 \%$), whereas the change in the radial electric field is significant ($E_r \! \left( \rho = 0.45 \right)_{max} - E_r \! \left( \rho = 0.45 \right)_{min} \simeq 6 \ \mathrm{kV} \! / \mathrm{m}$). Consequently, the dominant influence on fast-ion confinement will be that of the radial electric field.

\begin{figure*}[ht!]
    \centering
    \includegraphics[width=0.4\linewidth]{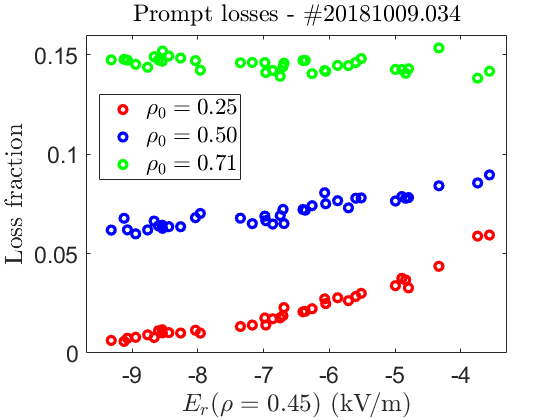}
    \caption{Energy prompt loss fraction as a function of $E_r$ at $\rho = 0.45$ for the selected plasmas.}
    \label{Fig_5_02_experimental_profiles_losses}
\end{figure*}

In figure \ref{Fig_5_02_experimental_profiles_losses} simulations show that for inner surfaces, $\rho_0 = 0.25$ and $\rho_0 = 0.50$, the losses decrease as the radial electric field increases (in absolute value). This is consistent with the previous theoretical considerations: a high radial electric field produces a high poloidal precession and this enhances fast-ion confinement. For the outer surface, $\rho_0 = 0.71$, there is no clear dependence of the losses. This is probably because $E_r$ profiles do not vary appreciably for $\rho > 0.7$. Moreover, the optimization of W7-X is focused on the core region \cite{drevlak2014fast}. Therefore, if a scan of this type is intended, it is crucial to generate the majority of ions close to the magnetic axis. In the rest of the work we will focus on these inner surfaces.

\begin{figure*}[ht!]
    \centering
    \includegraphics[width=0.4\linewidth]{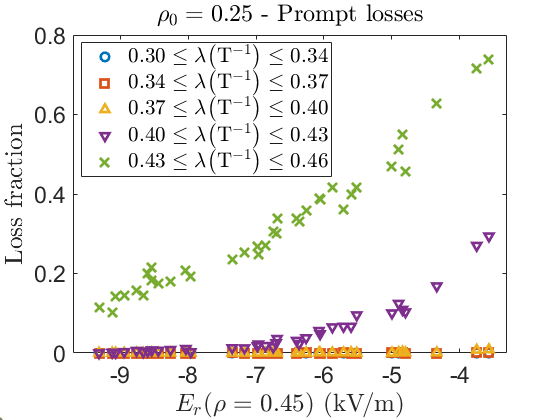}
    \hspace{0.5 cm}
    \includegraphics[width=0.4\linewidth]{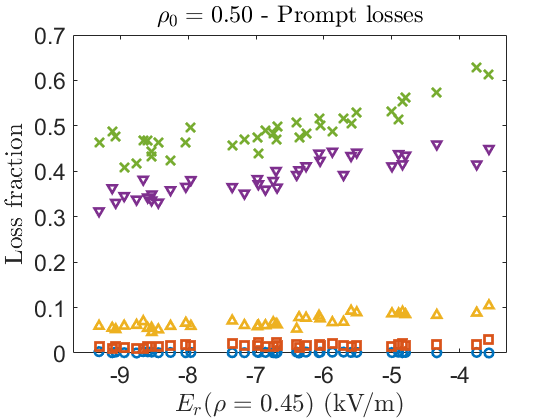}
    \caption{Energy prompt loss fraction as a function of $E_r$ at $\rho = 0.45$ for fast ions in different regions of the phase space for $\rho_0 = 0.25$ (left) and $\rho_0 = 0.50$ (right).}
    \label{Fig_5_03_losses_for_trapped}
\end{figure*}

Fast ions can be measured and/or generated in certain regions of the phase space. Thus we are going to study the dependence of fast-ion confinement on $E_r$ for different velocities. Trapped particles span a range of $\lambda$ that depends on the flux surface. However, for the high mirror configuration, the range does not vary significantly and is approximately $0.34 \ \mathrm{T}^{-1} \leq \lambda \leq 0.46 \ \mathrm{T}^{-1}$. The losses for fast ions in certain ranges of $\lambda$ are shown in figure \ref{Fig_5_03_losses_for_trapped}. We obtain analogous results for the different regions of the phase space studied in this analysis: the losses decrease as the radial electric field increases in absolute value. The only difference between the results for the different ranges in $\lambda$ is that this reduction in the losses is stronger for the more deeply-trapped particles (higher $\lambda$) and for the innermost surface ($\rho_0 = 0.25$).

We include one range in figure \ref{Fig_5_03_losses_for_trapped} that corresponds to passing particles (the one with $0.30 \ \mathrm{T}^{-1} \leq \lambda \leq 0.34 \ \mathrm{T}^{-1}$) and simulations show null losses for this range in $\lambda$. This is expected, as passing particles are confined, and collisions have not had enough time to turn them into trapped particles. In fact, simulations also show null losses for the most barely-trapped particles ($0.34 \ \mathrm{T}^{-1} \leq \lambda \leq 0.37 \ \mathrm{T}^{-1}$) born in the innermost surface. This is not surprising, since these particles are close to being passing, in the sense that they explore a large fraction of the flux surface and their radial drift almost averages to zero.

We now focus on trapped particles in the range $0.40 \ \mathrm{T}^{-1} \leq \lambda \leq 0.46 \ \mathrm{T}^{-1}$, born on the innermost surface. This way, they present a sufficiently strong dependence on $\left \langle \beta \right \rangle$ and $E_r$, while covering a relatively large range in $\lambda$. In figure \ref{Fig_5_04_losses_deeply_trapped_comp} it can be noticed that, when the ranges are properly chosen, the prompt losses from the experimentally-based scan decrease as the radial electric field increases (in absolute value) in a similar way to the decrease in the prompt losses from the ideal scan when $\left \langle \beta \right \rangle$ is increased. Specifically, for deeply-trapped ions and close to the axis, the scan on $E_r$, in the range $E_r \! \left( \rho = 0.45 \right) = \left[ -9 , -3 \right] \ \mathrm{kV} \! / \mathrm{m}$, during the discharge \#20181009.034 has the same effect as an ideal scan on $\left \langle \beta \right \rangle$ in the range $\left \langle \beta \right \rangle = \left[ 3.25 , 6 \right] \%$.

\section{Conclusions}
\label{conclusions}

In this work, fast-ion confinement in the high mirror configuration of W7-X has been characterized numerically for a variety of scenarios via the ASCOT5 code. Through an academic scan on $\left \langle \beta \right \rangle$ and $E_r$, we have shown that the region of maximal losses obtained with the simulations is compatible with the region of null bounce-average of the total tangential drift, showing a good agreement between simulations and analytical derivations. In addition, this academic scan shows that losses decrease monotonously when departing from this region of plasma parameters. This confirms the expectation that a large tangential drift enhances fast-ion confinement, which is one of the key points in W7-X optimization strategy regarding fast-ion confinement.

Specifically, via two ideal 1D scans on $\left \langle \beta \right \rangle$ and $E_r$ separately, the radial electric field effect on fast-ion confinement has been confirmed to be quantitatively equivalent to the one caused by $\left \langle \beta \right \rangle$. As a consequence, one could propose a scan on the radial electric field for a (partial) experimental validation of the W7-X optimization strategy, consisting of enhanced fast-ion confinement at large poloidal precession.

Through a preliminary scan on experimental profiles of W7-X, a possible scenario for this validation has been discussed. The requirements for a validation of this type have also been discussed: the variation in losses due to the increase of the tangential drift is significant only if fast ions are predominantly generated close to the magnetic axis and the trapped region of the phase space is sufficiently populated. As a consequence, if this scan is to be applied in the experiment, the heating scheme must be able to generate most of the fast ions under these conditions.

Future work should address the analysis of the possibility of generating and measuring fast ions in the phase space region where this effect is observed. This is not a trivial task since the current heating schemes of W7-X are far from meeting the aforementioned requirements.

Finally, the results presented in this work are relevant for any QI configuration that is proposed as a reactor candidate. Any intermediate-size device that is designed for the experimental validation of its confinement properties will face the same challenges regarding fast-ion confinement. A strategy similar to the one presented in this work will need to be implemented.

\section*{Acknowledgments}

The authors would like to thank the staff working at XULA (the High Performance Cluster at CIEMAT), the ASCOT team, the W7-X team and O.P. Ford for their valuable help throughout this work.

This work has been carried out within the framework of the EUROfusion Consortium, funded by the European Union via the Euratom Research and Training Programme (Grant Agreement No 101052200 -- EUROfusion). Views and opinions expressed are however those of the author(s) only and do not necessarily reflect those of the European Union or the European Commission. Neither the European Union nor the European Commission can be held responsible for them. This research was supported in part by Grants No. PID2021-123175NB-I00 and PID2024-155558OB-I00, funded by Ministerio de Ciencia, Innovación y Universidades / Agencia Estatal de Investigación / 10.13039/501100011033 and by ERDF/EU.

\appendix

\section{Calculation of $I_0$ and $I_{\left \langle \beta \right \rangle}$}
\label{calculation_I0_Ib}

In order to have an analytical estimate of the integrals $I_0$ and $I_{\left \langle \beta \right \rangle}$, we employed the procedure followed in \cite{velasco2023robust}. In that work, the magnetic field is assumed to be quasi-isodynamic (QI) and, using the formalism of the second adiabatic invariant, the equivalent to $\overline{\mathbf{v}_M \cdot \boldsymbol{\nabla} \alpha}$ is computed. Therefore, it is easy to extend those results to the integrals $I_0$ and $I_{\left \langle \beta \right \rangle}$, obtaining
\begin{equation} \label{eq_I_0_model}
    \small I_0 = \frac{\mu}{Z e \Psi_{LCFS}} \frac{\left( \partial_s B_{00}  + \partial_s \! \left| B_M \right| \right) \! K \! \left( \kappa^2 \right) - 2 \partial_s \! \left| B_M \right| \! E \! \left( \kappa^2 \right)}{K \! \left( \kappa^2 \right) + \frac{\left| B_M \right|}{B_{00}} \left( 2 E  \! \left( \kappa^2 \right) - K  \! \left( \kappa^2 \right) \right)}
\end{equation}
and
\begin{equation} \label{eq_I_b_model}
    \small I_{\left \langle \beta \right \rangle} = \frac{\mu B_{00}}{Z e \Psi_{LCFS}} \frac{K \! \left( \kappa^2 \right)}{K \! \left( \kappa^2 \right) + \frac{\left| B_M \right|}{B_{00}} \left( 2 E  \! \left( \kappa^2 \right) - K  \! \left( \kappa^2 \right) \right)} ,
\end{equation}
where
\begin{equation}
    K \! \left( \kappa^2 \right) = \int_{0}^{\pi / 2} \!\! \frac{1}{\sqrt{1 - \kappa^2 \sin^2{x}}} \ \mathrm{d} x
\end{equation}
is the complete elliptic integral of the first kind,
\begin{equation}
    E \! \left( \kappa^2 \right) = \int_{0}^{\pi / 2} \!\! \sqrt{1 - \kappa^2 \sin^2{x}} \ \mathrm{d} x
\end{equation}
is the complete elliptic integral of the second kind,
\begin{equation}
    \kappa^2 = \frac{1 - \lambda \! \left( B_{00} - \left| B_M \right|\right)}{2 \lambda \! \left| B_M \right|} ,
\end{equation}
$B_{00} \! \left( s \right)$ is the first term in the Fourier expansion of the magnetic field strength,
\begin{equation}
    B \! \left( s, \theta, \zeta \right) = \sum_{n,m} B_{mn} \! \left( s \right) \cos{\left( m \theta - N_{fp} n \zeta \right)} ,
\end{equation}
$N_{fp}$ is the number of toroidal field periods and
\begin{equation}
    B_M \! \left( s \right) = B \! \left( s, \theta, \zeta = 0 \right) - B_{00} \! \left( s \right) = \sum_{n > 0} B_{0n} \! \left( s \right)
\end{equation}
is the mirror term. Here, all the magnitudes have to be evaluated in vacuum. We have dropped all the ``$vac$" subscripts for simplicity.

\section*{References}

\bibliographystyle{ieeetr}
\bibliography{references}

@article{helander2012stellarator,
  title={Stellarator and tokamak plasmas: a comparison},
  author={Helander, Per and Beidler, C D and Bird, T M and Drevlak, M and Feng, Y and Hatzky, R and Jenko, F and Kleiber, R and Proll, J H E and Turkin, Yu and Xanthopoulos, P},
  journal={Plasma Physics and Controlled Fusion},
  volume={54},
  number={12},
  pages={124009},
  year={2012},
  publisher={IOP Publishing}
}

@article{helander2014theory,
  title={Theory of plasma confinement in non-axisymmetric magnetic fields},
  author={Helander, Per},
  journal={Reports on Progress in Physics},
  volume={77},
  number={8},
  pages={087001},
  year={2014},
  publisher={IOP Publishing}
}

@article{calvo2014optimizing,
  title={Optimizing stellarators for large flows},
  author={Calvo, Iv{\'a}n and Parra, Felix I and Alonso, Arturo and Velasco, Jos{\'e} Luis},
  journal={Plasma Physics and Controlled Fusion},
  volume={56},
  number={9},
  pages={094003},
  year={2014},
  publisher={IOP Publishing}
}

@article{plunk2019direct,
  title={Direct construction of optimized stellarator shapes. {P}art 3. {O}mnigenity near the magnetic axis},
  author={Plunk, Gabriel G and Landreman, Matt and Helander, Per},
  journal={Journal of Plasma Physics},
  volume={85},
  number={6},
  pages={905850602},
  year={2019},
  publisher={Cambridge University Press}
}

@article{landreman2022magnetic,
  title={Magnetic {F}ields with {P}recise {Q}uasisymmetry for {P}lasma {C}onfinement},
  author={Landreman, Matt and Paul, Elizabeth},
  journal={Physical Review Letters},
  volume={128},
  number={3},
  pages={035001},
  year={2022},
  publisher={APS}
}

@article{mata2022direct,
  title={Direct construction of stellarator-symmetric quasi-isodynamic magnetic configurations},
  author={Camacho Mata, Katia and Plunk, Gabriel G and Jorge, Rogerio},
  journal={Journal of Plasma Physics},
  volume={88},
  number={5},
  pages={905880503},
  year={2022},
  publisher={Cambridge University Press}
}

@article{landreman2022mapping,
  title={Mapping the space of quasisymmetric stellarators using optimized near-axis expansion},
  author={Landreman, Matt},
  journal={Journal of Plasma Physics},
  volume={88},
  number={6},
  pages={905880616},
  year={2022},
  publisher={Cambridge University Press}
}

@article{sanchez2023quasi,
  title={A quasi-isodynamic configuration with good confinement of fast ions at low plasma $\beta$},
  author={S{\'a}nchez, E and Velasco, J L and Calvo, I and Mulas, S},
  journal={Nuclear Fusion},
  volume={63},
  number={6},
  pages={066037},
  year={2023},
  publisher={IOP Publishing}
}

@article{goodman2023constructing,
  title={Constructing precisely quasi-isodynamic magnetic fields},
  author={Goodman, Alan G and Camacho Mata, K and Henneberg, Sophia A and Jorge, Rogerio and Landreman, Matt and Plunk, G G and Smith, H M and Mackenbach, R J J and Beidler, C D and Helander, P},
  journal={Journal of Plasma Physics},
  volume={89},
  number={5},
  pages={905890504},
  year={2023},
  publisher={Cambridge University Press}
}

@article{dudt2024magnetic,
  title={Magnetic fields with general omnigenity},
  author={Dudt, Daniel W and Goodman, Alan G and Conlin, Rory and Panici, Dario and Kolemen, Egemen},
  journal={Journal of Plasma Physics},
  volume={90},
  number={1},
  pages={905900120},
  year={2024},
  publisher={Cambridge University Press}
}

@article{garcia2024reduced,
  title={Reduced electrostatic turbulence in the quasi-isodynamic stellarator configuration {CIEMAT}-{QI}4},
  author={Garc{\'\i}a-Rega{\~n}a, J M and Calvo, I and S{\'a}nchez, E and Thienpondt, H and Velasco, J L and Capit{\'a}n, J A},
  journal={Nuclear Fusion},
  volume={65},
  number={1},
  pages={016036},
  year={2024},
  publisher={IOP Publishing}
}

@article{hall1975three,
  title={Three-dimensional equilibrium of the anisotropic, finite-pressure guiding-center plasma: {T}heory of the magnetic plasma},
  author={Hall, Laurence S and McNamara, Brendan},
  journal={The Physics of Fluids},
  volume={18},
  number={5},
  pages={552--565},
  year={1975},
  publisher={AIP Publishing}
}

@article{cary1997helical,
  title={Helical {P}lasma {C}onfinement {D}evices with {G}ood {C}onfinement {P}roperties},
  author={Cary, John R and Shasharina, Svetlana G},
  journal={Phys. Rev. Lett.},
  volume={78},
  number={4},
  pages={674},
  year={1997},
  publisher={APS}
}

@article{parra2015less,
  title={Less constrained omnigeneous stellarators},
  author={Parra, Felix I and Calvo, Iv{\'a}n and Helander, Per and Landreman, Matt},
  journal={Nuclear Fusion},
  volume={55},
  number={3},
  pages={033005},
  year={2015},
  publisher={IOP Publishing}
}

@article{velasco2024piecewise,
  title={Piecewise omnigenous stellarators},
  author={Velasco, J L and Calvo, I and Escoto, F J and S{\'a}nchez, E and Thienpondt, H and Parra, FI},
  journal={Physical Review Letters},
  volume={133},
  number={18},
  pages={185101},
  year={2024},
  publisher={APS}
}

@article{velasco2025exploration,
  title={Exploration of the parameter space of piecewise omnigenous stellarator magnetic fields},
  author={Velasco, J L and S{\'a}nchez, E and Calvo, I},
  journal={Nuclear Fusion},
  volume={65},
  number={5},
  pages={056012},
  year={2025},
  publisher={IOP Publishing}
}

@article{calvo2025new,
  title={A new class of optimized stellarators with zero bootstrap current},
  author={Calvo, Ivan and Velasco, Jose Luis and Helander, Per and Parra, Felix I},
  journal={arXiv preprint arXiv:2505.02546},
  year={2025}
}

@article{menon1981neutral,
  title={Neutral beam heating applications and development},
  author={Menon, Madhavan M},
  journal={Proceedings of the IEEE},
  volume={69},
  number={8},
  pages={1012--1029},
  year={1981},
  publisher={IEEE}
}

@article{stork1991neutral,
  title={Neutral beam heating and current drive systems},
  author={Stork, D},
  journal={Fusion Engineering and Design},
  volume={14},
  number={1-2},
  pages={111--133},
  year={1991},
  publisher={Elsevier}
}

@article{kirov2020synergistic,
  title={Synergistic {ICRH} and {NBI} heating for fast ion generation and maximising fusion rate in mixed plasmas at {JET}},
  author={Kirov, K K and Kazakov, Y and Nocente, M and Ongena, J and Baranov, Y and Casson, F and Eriksson, Jacob and Giacomelli, L and Hellesen, C and Kiptily, V and Bilato, R and others},
  journal={AIP Conference Proceedings},
  volume={2254},
  number={1},
  pages={030011},
  year={2020},
  publisher={AIP Publishing}
}

@article{rust2011w7xnbi,
  title={{W}7-{X} neutral-beam-injection: {S}election of the {NBI} source positions for experiment start-up},
  author={Rust, Norbert and Heinemann, Bernd and Mendelevitch, Boris and Peacock, Alan and Smirnow, Michael},
  journal={Fusion Engineering and Design},
  volume={86},
  number={6-8},
  pages={728--731},
  year={2011},
  publisher={Elsevier}
}

@article{machielsen2023fast,
  title={Fast ion generation by combined {RF}-{NBI} heating in {W}7-{X}},
  author={Machielsen, M and Graves, J P and Patten, H W and Slaby, C and Lazerson, S},
  journal={Journal of Plasma Physics},
  volume={89},
  number={2},
  pages={955890202},
  year={2023},
  publisher={Cambridge University Press}
}

@article{nuhrenberg1995overview,
  title={Overview on {W}endelstein 7-{X} theory},
  author={N{\"u}hrenberg, J and Lotz, W and Merkel, P and N{\"u}hrenberg, C and Schwenn, U and Strumberger, E and Hayashi, T},
  journal={Fusion Technology},
  volume={27},
  number={3T},
  pages={71--78},
  year={1995},
  publisher={Taylor \& Francis}
}

@article{klinger2016performance,
  title={Performance and properties of the first plasmas of {W}endelstein 7-{X}},
  author={Klinger, Thomas and Alonso, A and Bozhenkov, S and Burhenn, R and Dinklage, A and Fuchert, G and Geiger, J and Grulke, O and Langenberg, A and Hirsch, M and others},
  journal={Plasma Physics and Controlled Fusion},
  volume={59},
  number={1},
  pages={014018},
  year={2016},
  publisher={IOP Publishing}
}

@article{bosch2017final,
  title={Final integration, commissioning and start of the {W}endelstein 7-{X} stellarator operation},
  author={Bosch, H-S and Brakel, R and Braeuer, T and Bykov, V and van Eeten, P and Feist, J-H and F{\"u}llenbach, F and Gasparotto, M and Grote, H and Klinger, T and others},
  journal={Nuclear Fusion},
  volume={57},
  number={11},
  pages={116015},
  year={2017},
  publisher={IOP Publishing}
}

@article{wolf2017major,
  title={Major results from the first plasma campaign of the {W}endelstein 7-{X} stellarator},
  author={Wolf, R C and Ali, Anwaar and Alonso, Arturo and Baldzuhn, J{\"u}rgen and Beidler, Craig and Beurskens, Marc and Biedermann, C and Bosch, H-S and Bozhenkov, S and Brakel, Rudolf and Dinklage, A and others},
  journal={Nuclear Fusion},
  volume={57},
  number={10},
  pages={102020},
  year={2017},
  publisher={IOP Publishing}
}

@article{klinger2019overview,
  title={Overview of first {W}endelstein 7-{X} high-performance operation},
  author={Klinger, Thomas and Andreeva, T and Bozhenkov, Sergey and Brandt, C and Burhenn, R and Buttensch{\"o}n, Birger and Fuchert, G and Geiger, B and Grulke, O and Laqua, H P and Pablant, N and others},
  journal={Nuclear Fusion},
  volume={59},
  number={11},
  pages={112004},
  year={2019},
  publisher={IOP Publishing}
}

@article{beidler2021demonstration,
  title={Demonstration of reduced neoclassical energy transport in {W}endelstein 7-{X}},
  author={Beidler, C D and Smith, H M and Alonso, A and Andreeva, T and Baldzuhn, J and Beurskens, M N A and Borchardt, Matthias and Bozhenkov, S A and Brunner, Kai Jakob and Damm, Hannes and Drevlak, M and others},
  journal={Nature},
  volume={596},
  number={7871},
  pages={221--226},
  year={2021},
  publisher={Nature Publishing Group UK London}
}

@article{drevlak2014fast,
  title={Fast particle confinement with optimized coil currents in the {W}7-{X} stellarator},
  author={Drevlak, M and Geiger, J and Helander, P and Turkin, Yu},
  journal={Nuclear Fusion},
  volume={54},
  number={7},
  pages={073002},
  year={2014},
  publisher={IOP Publishing}
}

@article{wolf2018electron,
  title={Electron-cyclotron-resonance heating in {W}endelstein 7-{X}: {A} versatile heating and current-drive method and a tool for in-depth physics studies},
  author={Wolf, R C and Bozhenkov, S and Dinklage, A and Fuchert, G and Kazakov, Ye O and Laqua, H P and Marsen, S and Marushchenko, N B and Stange, T and Zanini, M and others},
  journal={Plasma Physics and Controlled Fusion},
  volume={61},
  number={1},
  pages={014037},
  year={2018},
  publisher={IOP Publishing}
}

@article{wobig1999theory,
  title={Theory of advanced stellarators},
  author={Wobig, Horst},
  journal={Plasma Physics and Controlled Fusion},
  volume={41},
  number={3A},
  pages={A159},
  year={1999},
  publisher={IOP Publishing}
}

@article{pablant2018core,
  title={Core radial electric field and transport in {W}endelstein 7-{X} plasmas},
  author={Pablant, N A and Langenberg, A and Alonso, A and Beidler, C D and Bitter, M and Bozhenkov, S and Burhenn, R and Beurskens, M and Delgado-Aparicio, L and Dinklage, A and others},
  journal={Physics of Plasmas},
  volume={25},
  number={2},
  pages={022508},
  year={2018},
  publisher={AIP Publishing}
}

@article{pablant2020investigation,
  title={Investigation of the neoclassical ambipolar electric field in ion-root plasmas on {W}7-{X}},
  author={Pablant, N A and Langenberg, Andreas and Alonso, A and Baldzuhn, J{\"u}rgen and Beidler, Craig D and Bozhenkov, S and Burhenn, Rainer and Brunner, Kai Jakob and Dinklage, Andreas and Fuchert, Golo and others},
  journal={Nuclear Fusion},
  volume={60},
  number={3},
  pages={036021},
  year={2020},
  publisher={IOP Publishing}
}

@article{carralero2020characterization,
  title={Characterization of the radial electric field and edge velocity shear in {W}endelstein 7-{X}},
  author={Carralero, D and Estrada, T and Windisch, T and Velasco, J L and Alonso, A and Beurskens, M and Bozhenkov, S and Damm, H and Fuchert, G and Gao, Y and others},
  journal={Nuclear Fusion},
  volume={60},
  number={10},
  pages={106019},
  year={2020},
  publisher={IOP Publishing}
}

@article{estrada2021radial,
  title={Radial electric field and density fluctuations measured by {D}oppler reflectometry during the post-pellet enhanced confinement phase in {W}7-{X}},
  author={Estrada, Teresa and Carralero, D and Windisch, T and S{\'a}nchez, E and Garc{\'\i}a-Rega{\~n}a, J M and Mart{\'\i}nez-Fern{\'a}ndez, J and De La Pe{\~n}a, A and Velasco, J L and Alonso, A and Beurskens, M and others},
  journal={Nuclear Fusion},
  volume={61},
  number={4},
  pages={046008},
  year={2021},
  publisher={IOP Publishing}
}

@article{carralero2021experimental,
  title={An experimental characterization of core turbulence regimes in {W}endelstein 7-{X}},
  author={Carralero, Daniel and Estrada, Teresa and Maragkoudakis, Emmanouil and Windisch, Thomas and Alonso, Arturo and Beurskens, M and Bozhenkov, S and Calvo, Ivan and Damm, Hannes and Ford, O P and others},
  journal={Nuclear Fusion},
  volume={61},
  number={9},
  pages={096015},
  year={2021},
  publisher={IOP Publishing}
}

@article{alonso2022plasma,
  title={Plasma flow measurements based on charge exchange recombination spectroscopy in the {W}endelstein 7-{X} stellarator},
  author={Alonso, A and Ford, O P and Van{\'o}, L and {\"A}k{\"a}slompolo, S and Buller, S and McDermott, R and Smith, H M and Baldzuhn, J and Beidler, C D and Beurskens, M and others},
  journal={Nuclear Fusion},
  volume={62},
  number={10},
  pages={106005},
  year={2022},
  publisher={IOP Publishing}
}

@article{kolesnichenko2006effects,
  title={Effects of the radial electric field on the confinement of trapped fast ions in the {W}endelstein 7-{X} and {H}elias reactor},
  author={Kolesnichenko, Ya I and Lutsenko, V V and Tykhyy, A V and Weller, A and Werner, A and Wobig, H and Geiger, J},
  journal={Physics of Plasmas},
  volume={13},
  number={7},
  pages={072504},
  year={2006},
  publisher={AIP Publishing}
}

@article{green2026energetic,
  title={Energetic particle loss under the effects of the radial electric field in {LHD}},
  author={Green, Ethan Maxwell and Hayashi, Wataru and Wei, Xishuo and Lin, Zhihong and Yamaguchi, Hiroyuki and Osakabe, Masaki and Nuga, Hideo and Ogawa, Kunihiro and Seki, Ryosuke and Isobe, Mitsutaka and Kawamoto, Yasuko and Shimizu, Akihiro and Ido, Takeshi and Nishiura, Masaki},
  journal={Nuclear Fusion},
  year={2026},
  publisher={IOP Publishing}
}

@article{lazerson2021modeling,
  title={Modeling and measurement of energetic particle slowing down in {W}endelstein 7-{X}},
  author={Lazerson, Samuel A and Pfefferl{\'e}, David and Drevlak, Michael and Smith, H{\aa}kan and Geiger, Joachim and {\"A}k{\"a}slompolo, Simppa and Xanthopoulos, Pavlos and Dinklage, Andreas and Ford, Oliver and McNeely, Paul and Rust, Norbert and Bozhenkov, Sergey and Hartmann, Dirk and Rahbarnia, Kian and Andreeva, Tamara and Schilling, Jonathan and Brandt, Christian and Neuner, Ulrich and Thomsen, Henning and Wolf, Robert C. and others},
  journal={Nuclear Fusion},
  volume={61},
  number={9},
  pages={096005},
  year={2021},
  publisher={IOP Publishing}
}

@article{faustin2016fast,
  title={Fast particle loss channels in {W}endelstein 7-{X}},
  author={Faustin, J M and Cooper, W A and Graves, J P and Pfefferl{\'e}, David and Geiger, Joachim},
  journal={Nuclear Fusion},
  volume={56},
  number={9},
  pages={092006},
  year={2016},
  publisher={IOP Publishing}
}

@article{hazeltine1973plasma,
  title={Plasma transport in a torus of arbitrary aspect ratio},
  author={Hazeltine, R D and Hinton, F L and Rosenbluth, M N},
  journal={The Physics of Fluids},
  volume={16},
  number={10},
  pages={1645--1653},
  year={1973},
  publisher={American Institute of Physics}
}

@article{velasco2021model,
  title={A model for the fast evaluation of prompt losses of energetic ions in stellarators},
  author={Velasco, J L and Calvo, I and Mulas, S and S{\'a}nchez, E and Parra, F I and Cappa, A and others},
  journal={Nuclear Fusion},
  volume={61},
  number={11},
  pages={116059},
  year={2021},
  publisher={IOP Publishing}
}

@article{nemov2008poloidal,
  title={Poloidal motion of trapped particle orbits in real-space coordinates},
  author={Nemov, V. V. and Kasilov, S. V. and Kernbichler, W. and Leitold, G. O.},
  journal={Physics of Plasmas},
  volume={15},
  number={5},
  pages={052501},
  year={2008},
  publisher={AIP Publishing}
}

@article{herbemont2022finite,
  title={Finite orbit width effects in large aspect ratio stellarators},
  author={d'Herbemont, Vincent and Parra, Felix I and Calvo, Iv{\'a}n and Velasco, Jos{\'e} Luis},
  journal={Journal of Plasma Physics},
  volume={88},
  number={5},
  pages={905880507},
  year={2022},
  publisher={Cambridge University Press}
}

@article{velasco2023robust,
  title={Robust stellarator optimization via flat mirror magnetic fields},
  author={Velasco, Jose L and Calvo, I and S{\'a}nchez, E and Parra, F I},
  journal={Nuclear Fusion},
  volume={63},
  number={12},
  pages={126038},
  year={2023},
  publisher={IOP Publishing}
}

@article{hirshman1983steepest,
  title={Steepest descent moment method for three-dimensional magnetohydrodynamic equilibria},
  author={Hirshman, Steven P and Whitson, J C},
  journal={The Physics of Fluids},
  volume={26},
  number={12},
  pages={3553--3568},
  year={1983},
  institution={Oak Ridge National Lab.(ORNL), Oak Ridge, TN (United States)}
}

@techreport{hirshman1985optimized,
  title={Optimized fourier representations for three-dimensional magnetic surfaces},
  author={Hirshman, S P and Meier, H K},
  journal={The Physics of Fluids},
  volume={28},
  number={5},
  pages={1387--1391},
  year={1985},
  institution={Oak Ridge National Lab.(ORNL), Oak Ridge, TN (United States)}
}

@article{lazerson2016verification,
  title={Verification of the ideal magnetohydrodynamic response at rational surfaces in the {VMEC} code},
  author={Lazerson, Samuel A and Loizu, Joaquim and Hirshman, Steven and Hudson, Stuart R},
  journal={Physics of Plasmas},
  volume={23},
  number={1},
  pages={012507},
  year={2016},
  publisher={AIP Publishing}
}

@article{landreman2014comparison,
  title={Comparison of particle trajectories and collision operators for collisional transport in nonaxisymmetric plasmas},
  author={Landreman, Matt and Smith, H{\aa}kan M and Moll{\'e}n, Albert and Helander, Per},
  journal={Physics of Plasmas},
  volume={21},
  number={4},
  pages={042503},
  year={2014},
  publisher={AIP Publishing}
}

@article{landreman2013new,
  title={New velocity-space discretization for continuum kinetic calculations and {F}okker--{P}lanck collisions},
  author={Landreman, Matt and Ernst, Darin R},
  journal={Journal of Computational Physics},
  volume={243},
  pages={130--150},
  year={2013},
  publisher={Elsevier}
}

@article{varje2019high,
  title={High-performance orbit-following code {ASCOT}5 for {M}onte {C}arlo simulations in fusion plasmas},
  author={Varje, Jari and S{\"a}rkim{\"a}ki, Konsta and Kontula, Joona and Ollus, Patrik and Kurki-Suonio, Taina and Snicker, Antti and Hirvijoki, Eero and {\"A}k{\"a}slompolo, Simppa},
  journal={arXiv preprint arXiv:1908.02482},
  year={2019}
}

@phdthesis{varje2022energetic,
  title={Energetic particles in reactor-relevant plasmas: modelling and validation},
  author={Varje, Jari},
  year={2022},
  school={School of Science - Aalto University}
}

@phdthesis{sarkimaki2019modelling,
  title={Modelling and understanding fast particle transport in non-axisymmetric tokamak plasmas},
  author={S{\"a}rkim{\"a}ki, Konsta},
  year={2019},
  school={School of Science - Aalto University}
}

@article{hirvijoki2014ascot,
  title={{ASCOT}: {S}olving the kinetic equation of minority particle species in tokamak plasmas},
  author={Hirvijoki, Eero and Asunta, Otto and Koskela, Tuomas and Kurki-Suonio, Taina and Miettunen, Juho and Sipil{\"a}, Seppo and Snicker, Antti and {\"A}k{\"a}slompolo, Simppa},
  journal={Computer Physics Communications},
  volume={185},
  number={4},
  pages={1310--1321},
  year={2014},
  publisher={Elsevier}
}

@article{drevlak2005pies,
  title={{PIES} free boundary stellarator equilibria with improved initial conditions},
  author={Drevlak, Michael and Monticello, D and Reiman, A},
  journal={Nuclear Fusion},
  volume={45},
  number={7},
  pages={731},
  year={2005},
  publisher={IOP Publishing}
}

@inproceedings{grulke2025overview,
  title={Overview of {W}endelstein 7-{X} high-performance operation},
  author={Grulke, O. and others},
  booktitle ={30th IAEA Fusion Energy Conference},
  month=oct,
  year={2025},
  address={Chengdu, China}
}

@article{ford2024turbulence,
  title={Turbulence-reduced high-performance scenarios in {W}endelstein 7-{X}},
  author={Ford, O P and Beurskens, M and Bozhenkov, S A and Lazerson, S and Van{\'o}, L and Alonso, A and Baldzuhn, J and Beidler, C D and Biedermann, C and Burhenn, R and others},
  journal={Nuclear Fusion},
  volume={64},
  number={8},
  pages={086067},
  year={2024},
  publisher={IOP Publishing}
}

\end{document}